\documentclass{iau}
\usepackage{graphicx}
\usepackage{natbib}
\usepackage{aas_macros}		
\usepackage{url}

\def\vect#1{\mathbf{#1}}											
\def\vectg#1{\boldsymbol{#1}}										

\def\m#1{\boldsymbol{\mathsf{#1}}}									
\def\bMod#1{\left| \, #1 \, \right|}											

\def\dl#1{\partial_{#1}} 											

\def\degrees{^{\circ}}

\title[Filaments, Walls and Pancakes.] 
{The structural elements of the cosmic web}

\author[Bernard J.T. Jones \& Rien van de Weygaert]   
{Bernard J.T. Jones$^1$
and 
Rien van de Weygaert$^2$
\thanks{This article reflects some of the work of the \textbf{Groningen Cosmic Web Group}, which has at various times included thesis work and subsequent papers written with Miguel Aragon-Calvo, Marius Cautun and Erwin Platen.}
}

\affiliation{$^1$Kapteyn Astronomical Institute, University of Groningen, The Netherlands  \\ 
email: {\tt jones@astro.rug.nl} \\[\affilskip]
$^2$
Kapteyn Astronomical Institute, University of Groningen, The Netherlands  \\ 
email: {\tt weygaert@astro.rug.nl}
}

\pubyear{2014}
\volume{308}  
\pagerange{119--126}
\jname{The Zeldovich Universe: Genesis and Growth of the Cosmic Web}
\editors{R. van de Weygaert, S.F. Shandarin, E. Saar \& J. Einasto, eds.}

\begin{document}

\maketitle

\begin{abstract}
In 1970 Zel'dovich published a far-reaching paper presenting a simple equation describing the nonlinear growth of primordial density inhomogeneities.  The equation was remarkably successful in explaining the large scale structure in the Universe that we observe: a Universe in which the structure appears to be delineated by filaments and clusters of galaxies surrounding huge void regions.  In order to concretise this impression it is necessary to define these structural elements through formal techniques with which we can compare the Zel'dovich model and N-body simulations with the observational data.  

We present an overview of recent efforts to identify voids, filaments and clusters in both the observed galaxy distribution and in numerical simulations of structure formation.  We focus, in particular, on methods that involve no fine-tuning of parameters and that handle scale dependence automatically.   It is important that these techniques should result in finding structures that relate directly to the dynamical mechanism of structure formation. 
 
\end{abstract}

\firstsection 

\section{A Short Historical Introduction}

In the early 1960's there was an ongoing debate between the proponents of two different cosmological models.  There was also a debate as to whether or not galaxy clusters were stable agglomerations of galaxies.  The first of these was to be resolved in 1965 with the discovery of the cosmic background radiation (CBR) by \citet{PenziasWilson:1965} and its interpretation by \citet{Dicke_etal:1965}  in terms of the relict radiation from a hot big bang origin of the Universe.  The second debate would take several decades to resolve in terms of the existence of non-luminous ``dark matter'' in the Universe.  Nonetheless, progress on describing the origin of cosmic structure started very soon after the CBR discovery with seminal papers by 
\citet{Peebles:1965}, \citet{SachsWolfe:1967}, \citet{Silk:1967}, \citet{Zeldovich:1970}, \citet{Sunyaev:1970}, \citet{PeeblesYu:1970}, among others.

\subsection{The Zel'dovich Legacy}
\citet{Zeldovich:1970} presented a simple model describing the non-linear evolution of structure in a co-expanding frame of reference described by a coordinate system $\{\vect{x} \}: \vect{x} = \vect{r} / a(t)$.  $a(t)$ is the usual cosmic expansion factor and $\vect{r}$ is the position of the particle in the expanding space. The Zel'dovich \textit{ansatz} is that a particle starting from initial position $\vect{x}(t_0) = \vect{q}$  would move ballistically according to
\begin{equation}
\vect{x}  = \vect{q} - D(t) \vect{v} (\vect{q}), \quad  \mathrm{with} \quad \vect{v} =  \frac{\partial{\Phi}}{\partial{q_j}} 
\label{eq:zeltrans}
\end{equation} 
for some scalar velocity potential $\Phi(\vect{q})$.  Here, $\vect{v} (\vect{q})$ is a pseudo-velocity vector with which the particle is projected from the point $\vect{q}$ and $D(t)$ is the so-called \textit{linear growth factor} that determines the time dependence of the particle motion.  $D(t)$ mimics the effects of gravity in such a way that the growth of small amplitude density fluctuations accords with linear perturbation theory \citep{Peebles:1965}
.  It is the separation of the displacement into a temporal component, $D(t)$, and a spatial component $\vect{v}(\vect{q})$, depending only on the initial particle positions that underlies the elegance and power of this model.

A small element of volume $d^3\vect{q}$ in the initial configuration is carried by equation \ref{eq:zeltrans} into a a small element $d^3\vect{x}$ by the flow.  If there is no flux of material into or out of these elemental volumes, the ratio of these two volumes gives the evolution of the density.  Likewise, the relative shear and rotation of the volumes describes the local kinematics.  This is expressed in terms of the \textit{Zel'dovich deformation tensor} $\m{Z} = \{ Z_{ij} \}$ which has components 
\begin{equation}
Z_{ij} = \frac{\partial{x_i}}{\partial{q_j}} 
\end{equation}
The density evolution in the physical space is
\begin{equation}
\frac{\rho(\vect{x}, t)}{\rho_0(t)} = \mathrm{det} \m{Z} = \bMod{\frac{\partial{x_i}}{\partial{q_j}}}
= \frac{1}{(1 - D(t) \gamma_1)(1 - D(t) \gamma_2)(1 - D(t) \gamma_3)} 
\label{eq:nonlindensity}
\end{equation}
where the $\gamma_i$ are the eigenvalues of $\m{Z}$, ordered so that $\gamma_1 \ge \gamma_2 \ge \gamma_3$.  

\begin{figure}[t]
	\centering
    \includegraphics[bb=0 0 794 803, width=8.0cm]{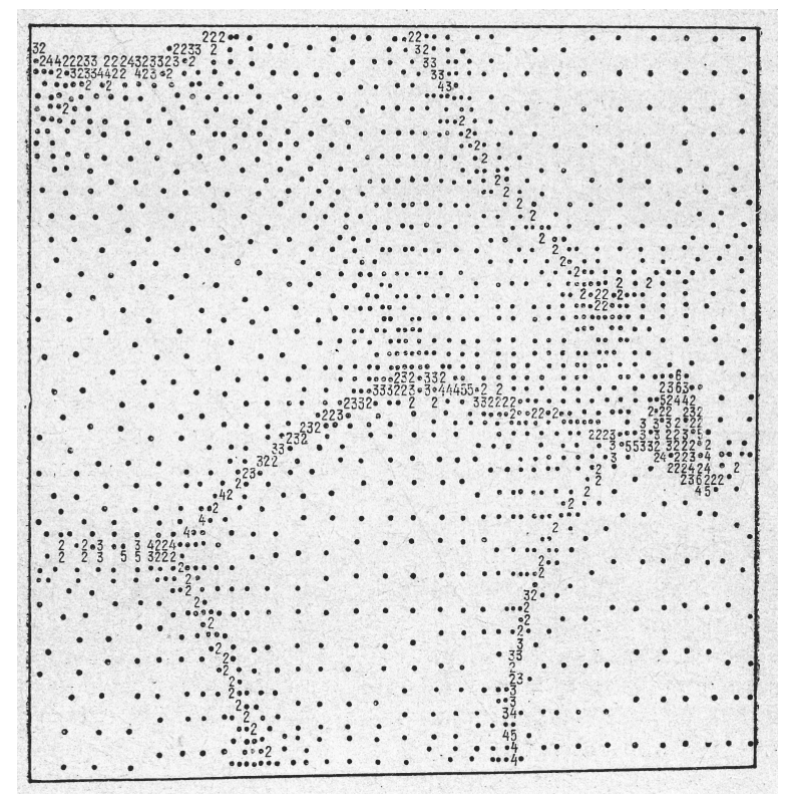}
	\caption{First 2-D simulation of the formation of structure using the Zel'dovich approximation \citep{DoroshShand:1978}.  This was done in 1976 at a time when filaments and voids had not been recognized as key elements in a web-like structure. The paper introduced use of the Hessian in defining local structural morphology.}
	\label{fig:shandarin}
\end{figure}

A key point about this approximation is that  the terms in the denominator all have the same time dependence: $D(t)$.  So the first term to become singular is the one involving the largest eigenvalue, $\gamma_1$.  Hence the first structure to form is was a flattened structure that became known as a \textit{pancake}.  This would be followed by the formation of a linear structure (a \textit{filament}) when the term involving $\gamma_2$ became singular and finally a point-like structure (a \textit{cluster}).  The pancakes would surround regions referred to as \textit{voids}.  The first numerical simulation of this process (Fig. \ref{fig:shandarin}) was due to  \citep{DoroshShand:1978}.  This was to be followed by more than a decade of increasingly sophisticated simulations of the Zel'dovich process.  

This was all done within the framework of a cosmological model dominated by hot dark matter in which the large scales would be the first to collapse.  Smaller scale structure would have to form via a fragmentation process taking place in the baryonic component.  The astrophysical consequences of equation \ref{eq:nonlindensity} were discussed at length by \citet{Doroshkevich:1974} and in terms of the gravitational and thermodynamic instability of the pancake structures by \citet{Jones:1981}. 

\subsection{Evidence for a cosmic web}
\begin{figure}[t]
	\centering
    \includegraphics[bb=0 0 1218 696, width=12.0cm]{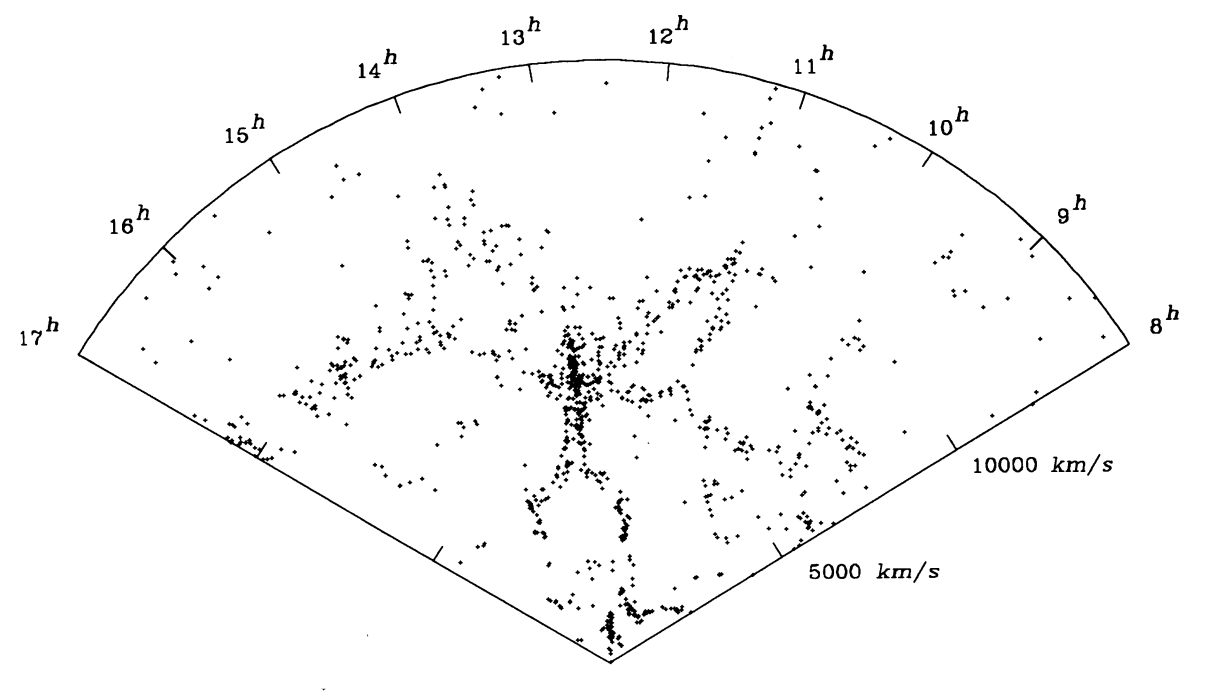}
	\caption{The CfA redshift slice \citet{deLapparent:1986}.  Although voids had earlier been recognized in surveys centered on rich galaxy clusters, this study demonstrated the ubiquity of voids.  Co-author John Huchra is famously quoted as saying \textit{``If we are right, these bubbles fill the universe just like suds filling the kitchen sink''} \citep{CfASliceNYT:1986}.
}
	\label{fig:delapparent}
\end{figure}

\begin{figure}[b]
	\centering
    \includegraphics[bb=0 0 2300 2200, width=12.0cm]{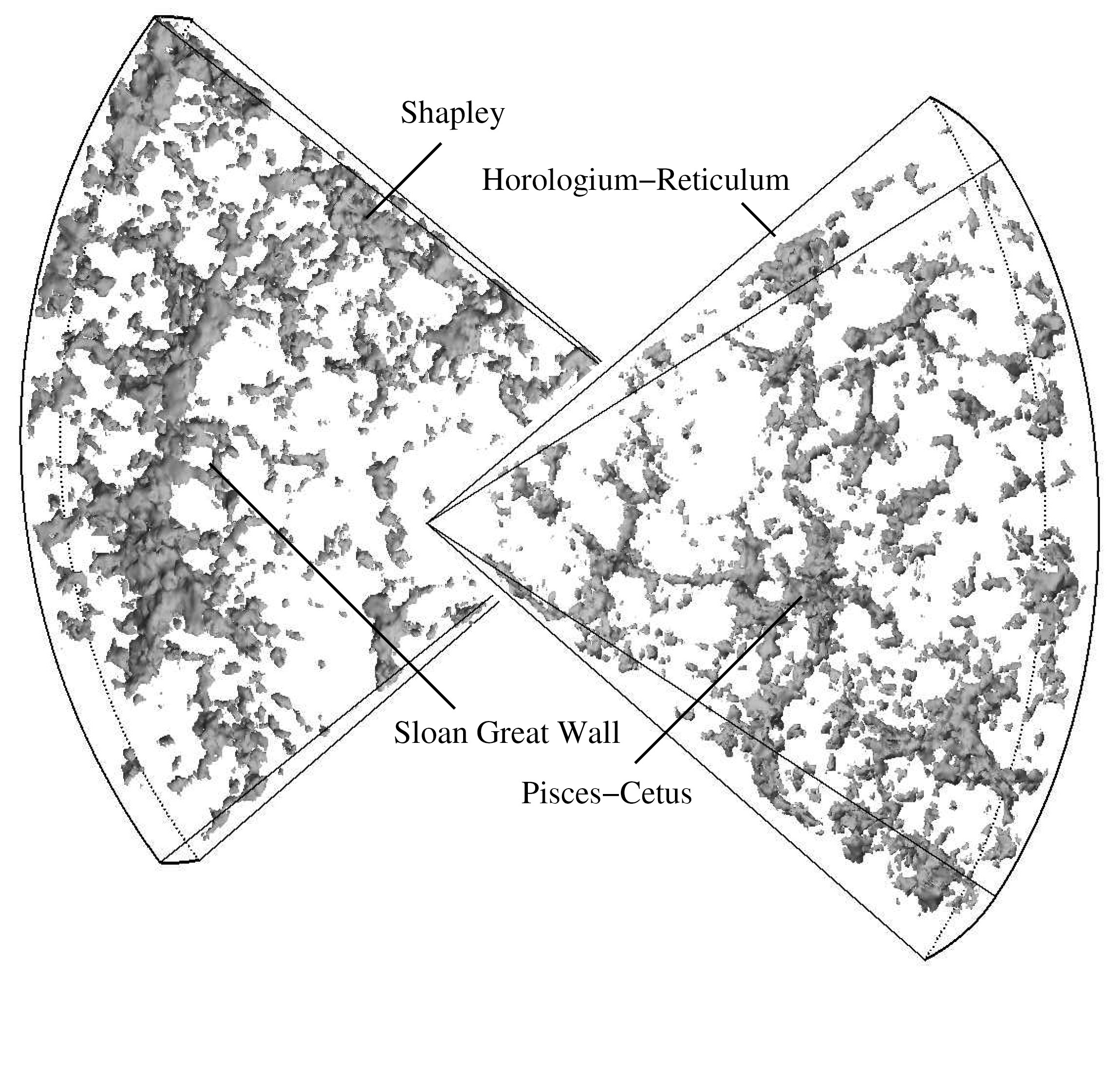}
	\caption{The distribution of galaxies in the 2dF Redshift Survey (2dFGRS) \citep{Colless:2001}.  The survey had redshifts for some 250,000 galaxies obtained with a 400-fibre multi-object spectrograph covering a $2\,\square\degrees$ field on the Anglo Australian Telescope.  The data was quickly made public, thereby enhancing the science value of the project. \newline
This smooth iso-density version of the survey map was produced from the original point distribution using the Delaunay Triangulation Field Estimator (`DTFE') of \citet{vdWSchaap:2007}.  DTFE provides an efficient mass-preserving mechanism for interpolating the field density at any point of the sample space.  Some of the most famous very large scale features are identified on the map.
}
	\label{fig:2dF}
\end{figure}

The first evidence for such large scale structures came from the seminal paper of \citet{deLapparent:1986} describing the CfA Redshift Slice (widely known as the \textit{deLapparent Slice}, see Figure \ref{fig:delapparent}).  
Previous to that others had identified voids by doing redshift surveys in the direction of rich galaxy clusters: in particular we think of the study of the Bo\"{o}tes void and of the Perseus-Pisces chain.  Subsequent ever larger redshift surveys served to enhance the first impression given by the de Lapparent slice.  In particular the \textit{2dF Redshift Survey} \citep{Colless:2001} not only showed the ubiquity of this structure (see Figure \ref{fig:2dF}), but also set the pattern in terms of the sample definition and analysis procedures for future redshift surveys.

\subsection{The cosmic web in numerical simulations}
One of the questions that arose in the late 1980's, with the popularization of the notion that the dark matter in the Universe might be cold \citep{Frenk:1985}, was whether such visually dominant structures could in fact arise in Cold Dark Matter (CDM) models where the structure was dominated by the very smallest scales.  \citet{DEFW:1985b} addressed this issue, at the same time introducing into simulations the important notion that the particles in the simulations that were to be identified with galaxies were only located where there were significant density peaks on galaxy scales 
\citep{Kaiser:1984, BBKS:1986}.  This was referred to as \textit{biasing}: it was a simple mechanism for identifying where the luminous matter might be found and for highlighting the structures we see.  With this mechanism, in those early models, the web-like structure was visible but not particularly striking.

\begin{figure}[t]
	\centering
    \includegraphics[bb=0 0 1070 803, width=13.5cm]{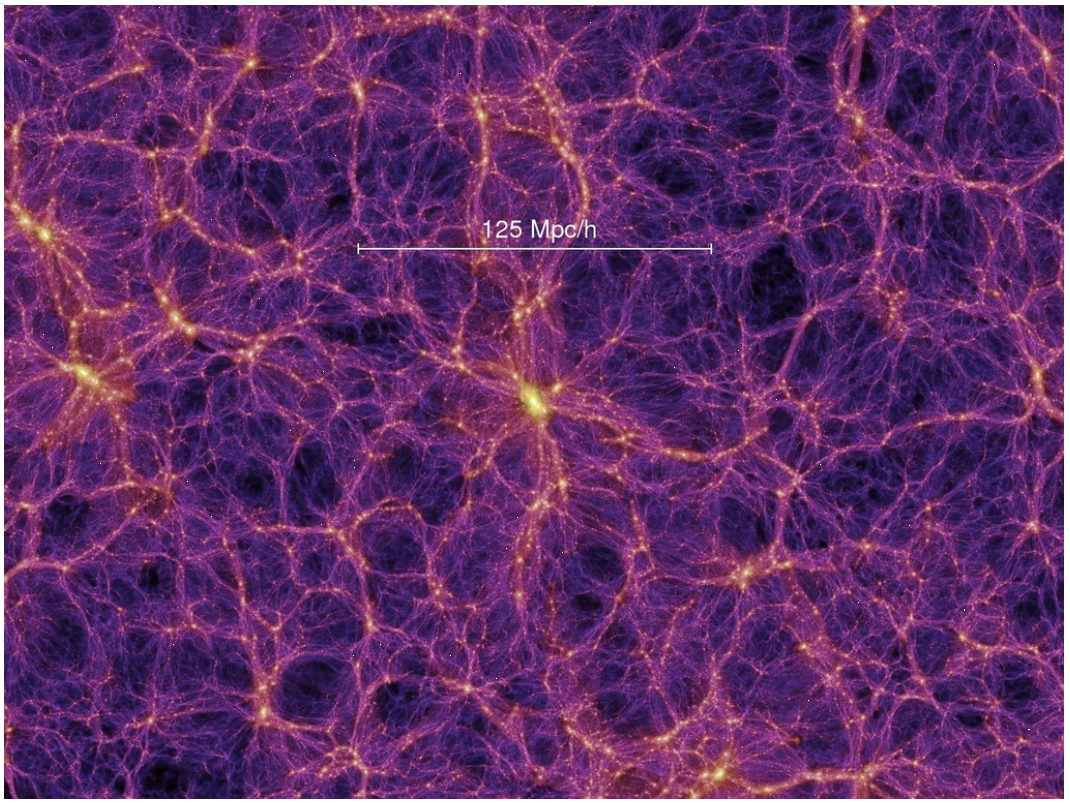}
	\caption{The cosmic web as seen through filamentary distribution of galaxies in the Virgo consortium's $10^{10}$-particle \textit{Millenium Simulation} \citep{Springel:2005}.  The subsequent decade has seen even larger simulations, some dealing with enhanced gas dynamic flow modeling (see the presentation of Pichon in these proceedings).
}
	\label{fig:Millenium}
\end{figure}

Larger simulations with improved models for identifying the sites of luminous galaxies did, however, bring the structure to light \citep[The \textit{Millenium Simulation}]{Springel:2005}.   Figure \ref{fig:Millenium} shows the structure in the galaxies that have been located in that simulation using semi-analytic models (`SAM') for galaxy evolution to identify haloes that host luminous galaxies. 

\section{Identification of Cosmic Structures}

The human brain is good at picking out structures from complex scenes, so we have little or no trouble in appreciating the presence of and identifying structural entities from pictures of galaxy distributions.  However, we guess, but do not know \textit{a priori}, that these structures arose as a consequence of the kind of dynamics suggested by the Zel'dovich model.  The issue is to relate what we perceive to dynamical models.  

To do that effectively we must have tools that will systematically and automatically isolate the structures that we perceive on the basis of point patterns taken either from galaxy redshift catalogs or numerical simulations.  With that we can then go further and discuss galaxy evolution in different environments.  We can also design observational projects to test our surmises about galaxy evolution in those various environments.

\subsection{Structure finding scenarios}
How you find structure in a distribution of particles depends on the information available in the source data.  There are the following possibilities to consider:
\begin{description}
\item [6-dimensional] \hfill \\ 
phase space in which both velocities and positions are specified for every particle.
\item [3-dimensional] \hfill \\ \indent
positions of all the particles
\item [2+1 dimensions] \hfill \\ 
2D positions on the sky + radial velocity
\end{description}

All of these three views are directly available from cosmological N-body models, whereas what we observe is only the ``2+1'' view.

It is not possible to review or even to list all published approaches to cosmological image segmentation in this short article.  However, many of these techniques are covered by other contributors to this volume.

\subsection{General Comments: Voids, Walls, Filaments and clusters}
The data we are presented with is almost always a set of points in one of the data-spaces described above.  The points are merely a statistical sample of the underlying structure that we wish to discover and analyze.  Most methods worked on a smoothed, grid-based, version of the given point set.  The idea of smoothing is that it somehow compensates or corrects for the statistical nature of the data we are presented with.  Smoothing of this underlying field was generally done with isotropic smoothing kernels.  This was fine for finding clusters of points, but vitiated against finding thin structures such as walls and filaments.  Nor was it very successful at finding voids, which, by definition, were regions where there were very few, if any, data points (see, for example, \citet{Colberg:2008}).  

The alternative to smoothing is re-sampling the given data onto a grid.  One of the central tools to do this was developed by  \citet{vdWSchaap:2007}: the Delaunay Triangulation Field Estimator (``DTFE'') or its dual, the Voronoi Tesseleation Field Estimator (``VTFE'').  The DTFE mechanism provides a continuous field on a grid that is free of structural bias and closely reflects the original point data.  Any noise present in the original data is still present in the gridded data, and so there is no cost resulting from assumptions about either the noise or the underlying data.

\begin{figure}[t]
	\centering
    \includegraphics[bb=0 0 1167 520, width=12.0cm]{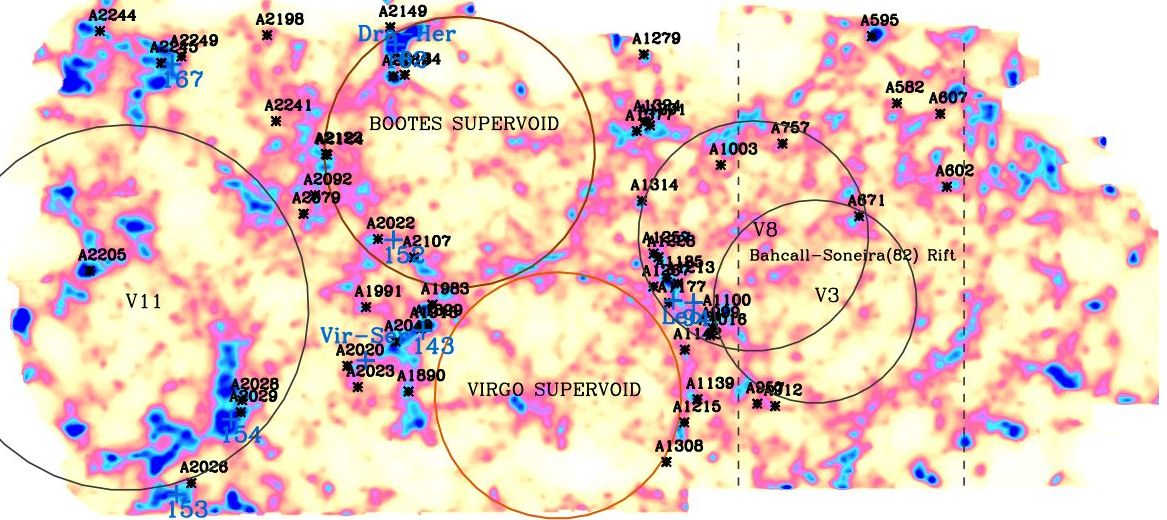}
	\caption{Super-voids found in the DR7 release of the Sloan Digital Sky Survey (SDSS) using the Watershed Void Finder of \citet{Platen:2007, Platen:2008}.  The super-void locations are depicted by a circle having a radius equal to the effective void radius.  The boundaries of the super-voids are in places well defined by aggregates (super-clusters) of rich galaxy clusters which are marked on the figure by black crosses.\newline  
The `Virgo super-void' is so-called because it lies in the direction of and far behind the Virgo cluster at a distance of $\sim140 h^{-1}$Mpc.  This is Void 7 in Table 1 of \citet{Tully:1986}, the other V-numbers are from that table.  These super-voids voids range in diameter assigned by the WVF method from $\sim 82h^{-1}$Mpc for V3 to $\sim 148h^{-1}$Mpc for V11.\newline
Sizes are typically $\sim 100h^{-1}$Mpc, which tempts us to identify them with the BAO scales.
}
	\label{fig:MSupervoids}
\end{figure}

\section{Void Finders}
The best way of finding voids appears to be the Watershed based methods, as first introduced by \citet[The Watershed Void Finder, WVF]{Platen:2007, Platen:2008} , and \citet[ZOBOV]{Neyrinck:2008}.    
\footnote{For an excellent and recent review of watershed and other void finding methods see \citet{Way:2014}.  See also the Void Finder comparison test of \citet{Colberg:2008}.}

The watershed method is parameter-free and has been widely used as the basis for identifying the other cosmic structures: walls, filaments and clusters.  Since the voids are thought to be space-filling, their intersections can be identified as walls, while the intersections of the walls are the filaments, and the meeting points of the filaments are the clusters: in fact, just as envisaged in the original pancake theory \citep{Doroshkevich:1974}.  

An example of this is the \textit{Spineweb} method of \citet{Aragon-Calvo:2010}.  However, one of the practical disadvantages of this approach is that filaments necessarily connect clusters, whereas in the real world we see filaments that emanate from clusters to terminate inside a void.  We also appear to see filaments within the walls \citep{Rieder:2013}. 

\subsection{WVF: Watershed Void Finder}
The Watershed Void Finder (WVF) of \citet{Platen:2007, Platen:2008} has a data preprocessing stage which involves re-sampling data onto a grid using the DTFE, and then smoothing the sampled data on different scales $R_f$ ranging from $1h^{-1}$ Mpc  to $10h^{-1}$.  Importantly, the data is then median filtered several times to remove small scale fluctuations that may be due to remaining smaller scale structure than the filter $R_f$.  This yields a clear basin and well defined surrounding edges for the watershed void finding.  This process makes it possible to find very large scale `voids' that are below-average in density while having smaller scale substructure. 

Examples of such super-voids found using WVF with a filter scale of $R_f = 10 h^{-1}$Mpc are shown in figure \ref{fig:MSupervoids}.  The WVF method recovers several super-voids that had been previously identified in the sample volume, eg: \citet{Tully:1986}.\footnote{
The diameters assigned by WVF are somewhat smaller than those given by Tully (renormalized to $H_0 = 100$).  Moreover, they are far from being spherical which makes assignment of a size somewhat ambiguous.}

\section{MMF-NEXUS: Multi-scale parameter-free analysis}
The idea of doing image segmentation by mimicking the brain goes back the the work of \citet{Marr:1982}  and his concept of the \textit{primal sketch}.  In a 2-dimensional continuous grey-tone image, Marr considered the contours on where the grey-scale level $\phi$ has $\nabla^2\phi = 0$.  By looking at differently (Gaussian) smoothed versions of the same image Marr constructed what he described as a \textit{primal sketch}.  Over the past decades this notion has been generalised and refined by a number of authors in diverse fields: one of the main changes has been to study the Hessian of the field $\phi$: ${\cal{H}}_{ab} = \partial^2 \phi/\partial x_a \partial x_b$.  This was further modified and brought to the problem of cosmic structure finding by \citet[and references therein]{Aragon-Calvo:2007b}. 

\subsection{Halo spins and bulk motions}
\begin{figure}[t]
	\centering
    \includegraphics[bb=0 0 1630 832, width=13.5cm]{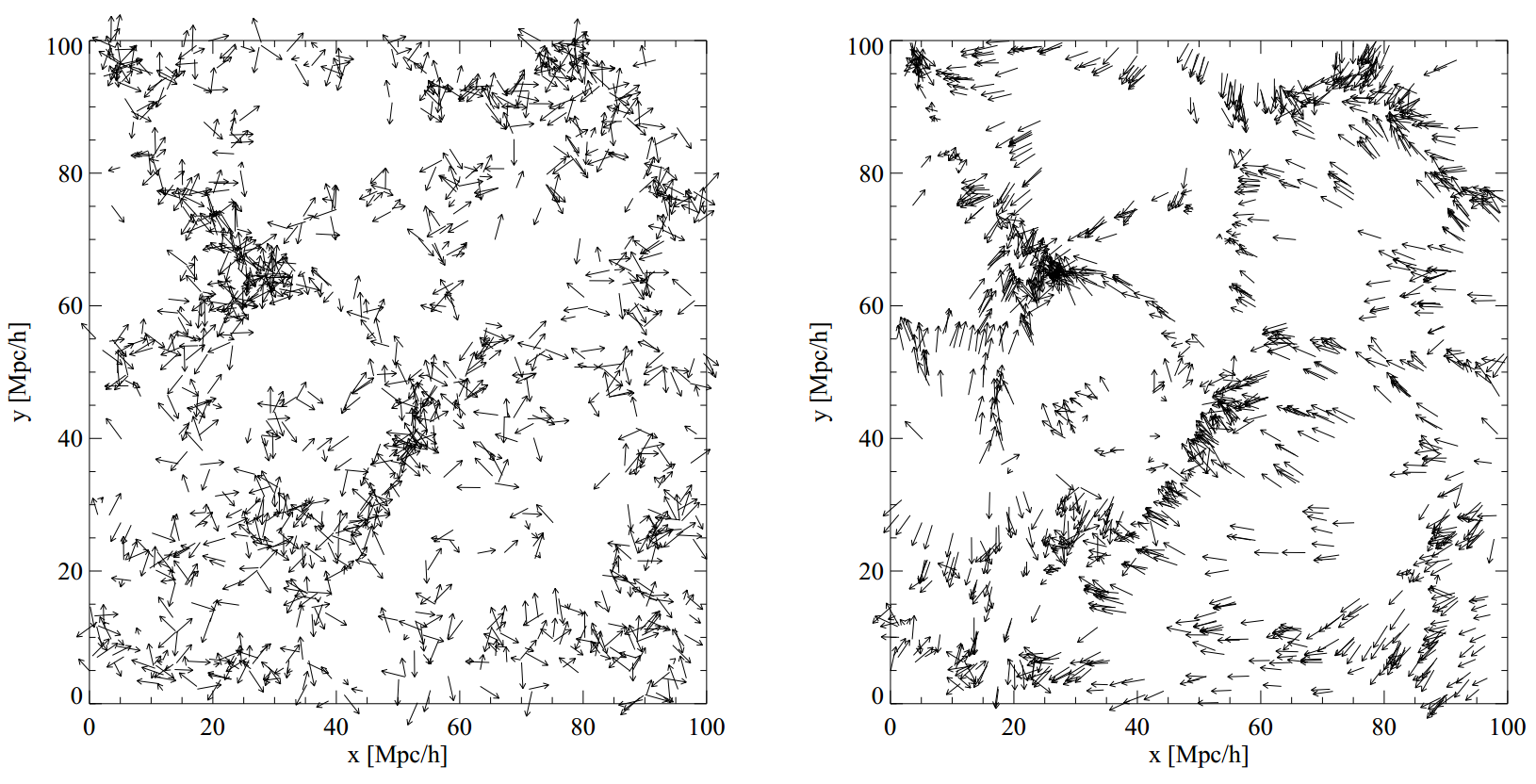}
	\caption{\textit{Left:} Orientation of halo spin directions.  \textit{Right:} Halo bulk motions.  The spin axes appear random, any systemic orientation is marginal.  However, the systemic bulk flows of filaments is manifest.  (From numerical simulations reported by \citet{Trowland:2013}).
 }
	\label{fig:Trowland}
\end{figure}
One of the more open questions of cosmology has always been the question of whether there are any systematic alignments in the spins of galaxies.  The advent of large N-body simulations has allowed this question to be studied within the framework of the dark matter halos that form in these simulations.  Use of simulations greatly simplifies the question since there are no projection effects to deal with and the spin axes pf the halos is well defined.  Once we can ambiguously identify filaments, the issue is further simplified when fixing attention to looking for any alignment of galaxies in the filaments.  

The galaxies in the filaments are moving along the filaments under the influence of the general tidal field of the nearest galaxy cluster into which they are falling.  Hence one might expected that Filaments are a good place to search for alignments with the structure.  Moreover, since these are the same tidal fields that determine the bulk motions within the filaments, we would likewise expect to find correlations of spin axis with various aspects of the bulk flow.

\subsection{Alignments of halo spins and filaments}
The situation is nevertheless not totally straightforward, as can be seen from the left hand panel of Fig. \ref{fig:Trowland} \citep{Trowland:2013}: the small arrows depict the spin axes of haloes in the simulation and they are not manifestly lined up.  However, the right hand panel shows that the filaments do have a unambiguous bulk motions (see also \citet{Rieder:2013}).  

The first report of systemic alignment of spins in N-body filaments and walls was that of \citet{Aragon-Calvo:2007a} who did find evidence for non-random alignment in the halos, and, interestingly, reported that the low-mass objects tended to have there spin axes aligned within the host structure, while the higher mass objects had their spin axes aligned perpendicular to the host structure.   This result that has since been confirmed by a number of groups (see, for example, \citet{Forero-Romero:2014} who provide a nice overview of previous results).   There have also been confirmations that the filaments are correlated with various aspects the underlying velocity field \citep{WeyKamp:1993, Libeskind:2014, Temple:2014}. As seen in the right hand panel of Fig. \ref{fig:Trowland}, the correlations in the velocity field are rather strong. 

\subsection{Filamentary structure and Galaxy alignments in the SDSS}
The situation with real data ($2+1$ data) is less straightforward.  Redshift distortions make it more difficult to define filaments that are oriented close to the line of sight, this is evident in Figure \ref{fig:delapparent}.  Moreover, there is a difficulty in accurately defining the spin axis of real galaxies which are determined from their baryonic component rather than their dark matter component.  The selection of edge-on galaxies removes the spin axis orientation issue, and selection of well defined filaments that are almost transverse to the line of sight removes filaments that result from redshift distortions.  Using a sample of $\sim$500 filaments found in the DR5 release of the SDSS using MMF, \citet{Jones:2010} found evidence of a bimodal distribution of the spins relative to the local filament axis.  \citet{Temple:2013} analyzed filaments found in the SDSS (DR8) using their BISOUS method and found evidence for alignments in spiral galaxies with the spin axis along the filament, and for E/S0 galaxies, but with their minor axes aligned perpendicular to the host filament.

\section{NEXUS and NEXUS+}
\begin{figure}[t]
	\centering
    \includegraphics[bb=0 0 1070 682, width=13.5cm]{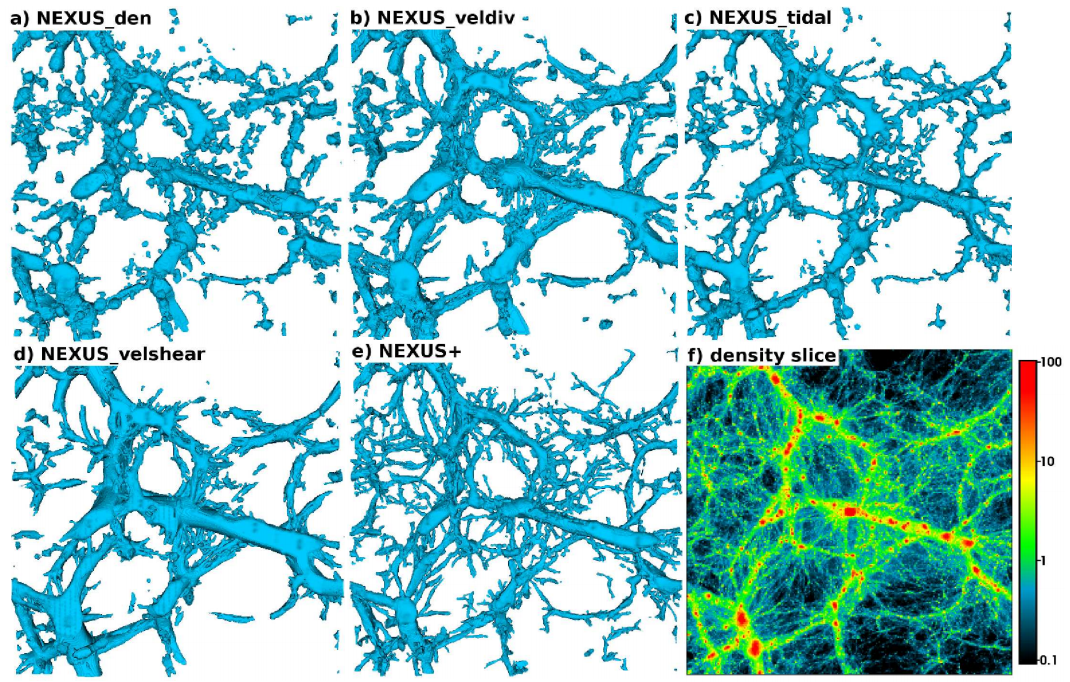}
	\caption{Structures in different fields as revealed by the NEXUS and NEXUS+ techniques. We clearly see the structural differences and inter-relationships between the various field that determine the physical nature of the cosmic web. The density field as delineated by NEXUS+ using the logarithm of the density shows a finer structure than using NEXUS on the density field.}
	\label{fig:NEXUS}
\end{figure}

NEXUS \citep{Cautun:2013} is a scale-free and parameter-free structure classifier and extends and improves on the original MMF structure classifier \citep{Aragon-Calvo:2007b}.  Like MMF, NEXUS was designed to handle density fields, but, importantly, it is equally aimed at determining the structural morphology of the divergence and shear of the particle velocity field, as well as the structure in the gravitational tidal field.   The NEXUS+ variant of NEXUS deals with the log-density field.  Structures found in the various fields are shown in Figure \ref{fig:NEXUS}. 

\subsection{Determining scale-free parameter-free morphology}
The field values are re-sampled onto a grid using DTFE, and then a hierarchy of smoothed fields is generated on a set of scales $\{k\}$ \footnote{The smoothing kernel used was Gaussian on a set of scales differing by factors of 2.  Earlier experiments by BJ (unpublished) had shown that the use of spline or wavelet kernels, or the use of finer hierarchies made relatively little difference when structural morphology was based simply on the eigenvalues of the Hessian of the field values.}   The local environment is characterized on each scale, $k$, by an \textit{environment signature}, $S_k$, that is constructed from the local eigenvalues of the Hessian of the underlying field values smoothed on that scale.  The scale free value of the environment signature, $S$, is then simply the maximum value of $S_k$.  

The map of $S$-values has a large dynamic range and only the largest value of $S$ are significant, the low ones being noise.  A threshold, $S_{thresh}$ must therefore be applied for each of the cluster, filament and wall-like structure maps: the significant structure is set of above-threshold points.   The major technical issue is the method whereby the critical threshold is determined, and how to determine that automatically.  Details are in \citet[Section 2]{Cautun:2013}.

One of the key technical differences between NEXUS and MMF lies in this automated thresholding mechanism.  For finding filamentary structure in the density field, MMF calculated a threshold by finding a critical value of a percolation parameter.  NEXUS uses a threshold value where the mass fraction, $\Delta M^2$, in the filaments is most rapidly changing. As it turns out, for the density field, the percolation and mass fraction thresholds are very close, and hence the resulting filamentary structure is very similar in both approaches.  Using $\Delta M^2$ solves an issue regarding the walls in that were found in MMF, and works particularly well for the structures based on properties of the velocity field or gravitational potential.  The details are given in \citet[Appendix A]{Cautun:2013}.

\subsection{NEXUS structures in N-body simulations}
Application of NEXUS+ to an N-body model, where the densities and velocities are known, allows structural characterization of the density field, the velocity divergence and shear fields, the tidal field and the log-density field.  Figure \ref{fig:NEXUS} displays the results. 

Although the perceived structure is similar for all tracers, the details of the structures that are found vary perceptibly among the different tracers used in the mapping.  For example, the filamentary structure found log-density field, labeled ``NEXUS+'' in the figure, covers a large range of scales: it is clearly advantageous to use this method for density fields.  The velocity field tracers tend to be smoother, but neverthelss show structures of all scales.  It is also to be noted that many filaments are often ``open-ended'' - they disappear into voids.  These differences are potentially important since they provide clues as to how the structures evolved.

\begin{figure}[t]
	\centering
    \includegraphics[bb=0 0 2800 900, width=13.5cm]{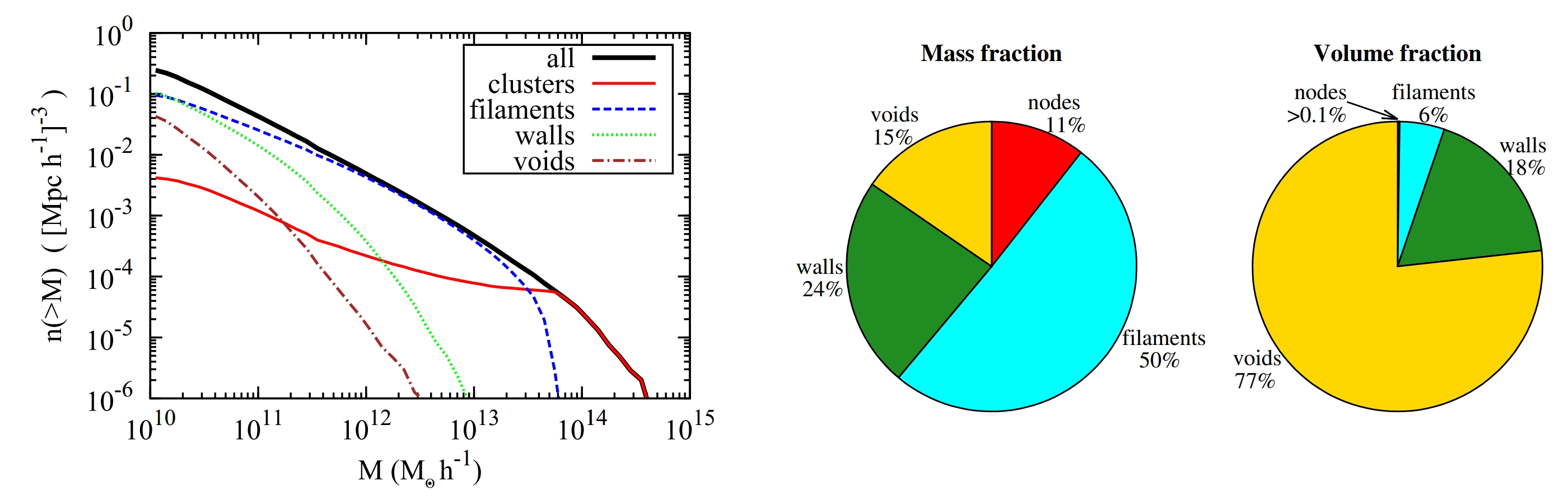}
	\caption{Left: The cumulative halo mass function segmented according to the components of the cosmic web as identified by NEXUS+, normalized according to the volume of the whole simulation box.  Right: The mass and volume fractions occupied by cosmic web environments detected by the NEXUS+ method.
	\citep[Figures 19, 8]{Cautun:2014}.
}
	\label{fig:NEXUS_Fractions}
\end{figure}

Galaxy formation and evolution are strongly environment dependent and NEXUS provides a tool for objectively identifying and analyzing the local environment of a galaxy at any stage of its evolution.  This is discussed exhaustively in \citet{Cautun:2014}, where the focus in on the halos in the simulation, and their morphological environments.  The simulations used are the $z=0$ volumes of the  MS-I \citep{Springel:2005} and MS-II \citep{Boylan-Kolchin:2009} simulations, which use the canonical WMAP cosmological parameters.  We show just one set of results from the NEXUS+ analysis analysis of these simulations in \ref{fig:NEXUS_Fractions}.   The graph in the left panel shows the cumulative mass function of objects as a function of environment. The right panel shows pie-diagrams of the morphological make-up by volume fraction and by mass fraction: the voids obviously dominate the volume fraction while the filaments dominate the mass fraction.

\section{The 6-dimensional and 3-dimensional views}
It is clear that $6$-dimensional, phase space, structure classifiers should do a better job at assigning cosmic structures to different morphology classes than do the $3$-dimensional classifiers or the $2+1$-dimensional classifiers.  However, the central issue is to understand what each of the classifiers is measuring and to discuss how the findings of these classifiers relate to one another.  As a specific example, consider the ORIGAMI and NEXUS methods.

\subsection{ORIGAMI and NEXUS}
The ORIGAMI approach \citep{Falck:2012, Neyrinck:2012} considers the full $6$-dimensional phase space structure at the time the initial conditions are set. The inter-penetration of opposingly moving streams of particles in $(x,y,z)$ configuration space appears as a folding of the phase space sheet on which the particle lie in the full $6$-dimensional phase space (see also \citet{Hidding:2014}).  The structure is delineated by these folds, hence the name ``origami''.  

NEXUS works entirely on the snapshot of the $3$-space particle configuration at any given time, examining the eigenvalues of the scale-independent Hessian of the density or log-density field, and takes no account of how that structure originated.  NEXUS is a powerful perceptual model having no dynamical rationale.  

\begin{figure}[t]
	\centering
    \includegraphics[bb=0 0 2000 647, width=\textwidth]{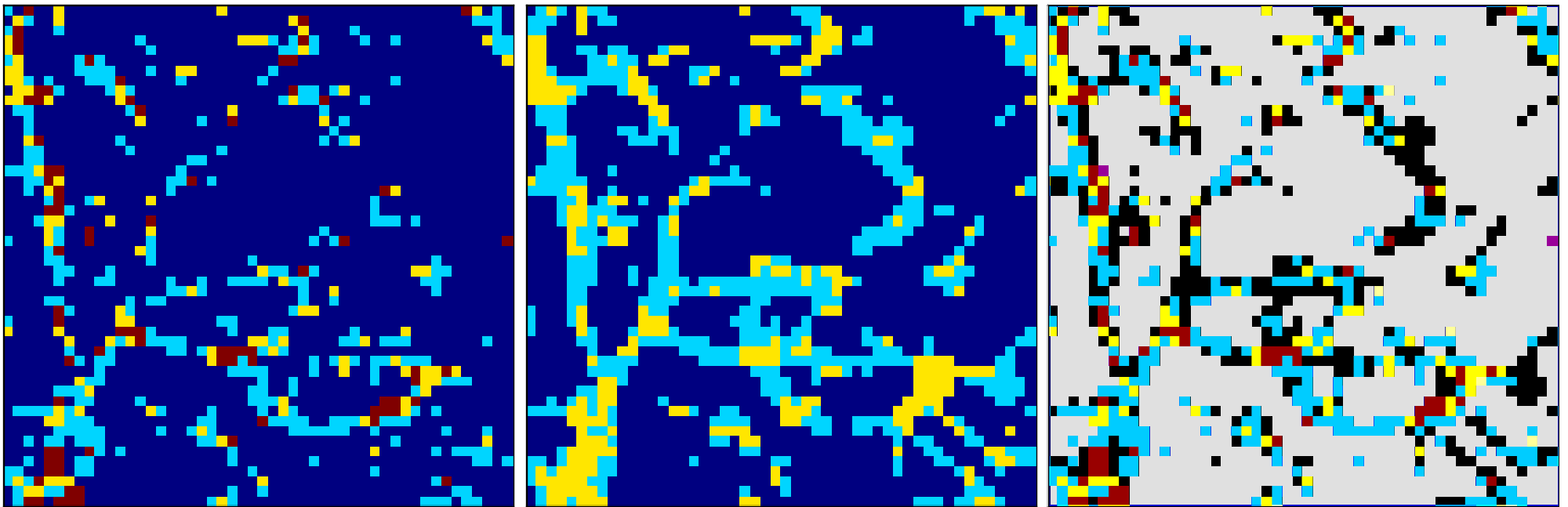}
	\caption{A simple comparison of the spatial structure found by the $6-D$ Origami and $3D$ Nexus filament finders.  The panel on the left is the outcome of the ORIGAMI method, while the center panel is the result from NEXUS.  The panel on the right has been constructed by recoloring all the NEXUS structure in black and transparently superposing the ORIGAMI image. Note that, in this, way some of the NEXUS (black) structures are hidden beneath the ORIGAMI structures and the gray area show where both agree there is no structure. (From the Lorentz Center Web Comparison Project, Leiden February 2014.)
}
	\label{fig:NEXUSORIGAMI}
\end{figure}
\subsection{Why do the ORIGAMI and NEXUS results look so similar?}
The images in Figure \ref{fig:NEXUSORIGAMI} show a slice from the results of ORIGAMI and NEXUS analysis of an N-body model that was provided to participants in the Lorentz Center Cosmic Web Comparison meeting (Leiden, February 2014) for the purpose of structure classification.  The spatial structures identified in the slice are highlighted\footnote{
Since this is a thin slice some of the apparently filamentary structure will in fact be a slice through a wall. 
}.  

The two left-most panels show the structure assignment by the ORIGAMI and NEXUS classifications schemes.  The rightmost panel highlights the commonality between the two approaches by showing the ORIGAMI identified structures superposed on the NEXUS image.  The NEXUS structure has been entirely rendered in black to make the comparison clearer.  This could be done with far more rigor, but even at this simple level we see the remarkable agreement between the two approaches.

The approaches are totally different, and yet the correspondence between the classifications, at first sight, appears quite remarkable.  We see something similar in Figure \ref{fig:NEXUS} where NEXUS analysis of an N-body simulation leads to density field maps (a) and (e) that compare in detail with the velocity divergence field and shear field maps (b) and (d).  This is, of course, a consequence of the equations of motion of the inhomogeneous cosmic medium: the Euler equation tells us that the rate of change of the velocity is governed by the gravitational potential gradient.  The velocity field gradient, ie. the velocity divergence and shear fields, are driven by the gravitational tidal shear, which, via the Poisson equation, is directly proportional to the density.  

In the Zel'dovich approximation this is almost trivial since in that approximation the velocity potential and the gravitational potential are proportional to one another.\footnote{
For flow having zero vorticity, the velocity field can be written $\vect{v} \propto \vectg{\nabla} \cal{U}$, where $\cal{U}$ is the velocity potential and the proportionality factor depends only on time.  If $\phi$ is the gravitational potential, then the Zel'dovich approximation is $\cal{U} \propto \phi$ \citep{Gramman:1993, Jones:1999}.  With this $\dl{i}\dl{j} \cal{U} = \dl{i}\dl{j} \phi $, where $\dl{i}$ denotes $\partial /\partial x_i$.  All the derivatives $\dl{i}\dl{j} \phi $ are determined by the density field (Poisson's equation and the boundary conditions).  $\dl{i}\dl{j} {\cal{U}} = \dl{i} v_j$ is the velocity shear tensor.
}  
\section{Concluding remarks}
Zeldovich's simple nonlinear approximation to the evolution of cosmic structure from given initial conditions is a powerful tool for understanding the evolution of the cosmic web.  The large scale structures that are the consequence of that approximation have a direct reflection in the web-like structures seen in N-body simulations.  

Objectively segmenting cosmic structure into its main component parts is a key factor in understanding the environmental issues which affect our understanding of galaxy evolution and out ability to analyses and interpret velocity fields.  There is now a wide variety of techniques available that achieve this segmentation, though most of them are not scale-independent and require tuning of parameters to obtain optimal results.  We have focused on our NEXUS/NEXUS+ scheme and shown, in that case, that there is a remarkable level of agreement between the structure found by ORIGAMI analysis of the Zel'dovich caustics in the phase space of the initial conditions and by NEXUS analysis of the final conditions. 
\section*{Acknowledgements}
BJ would like to thank the IAU and the organisers for making it possible to participate in this meeting.

\end{document}